\documentclass[11pt,a4paper,preprintnumbers]{article}
\pdfoutput=1
\usepackage{jheppub}
\usepackage{rotating}
\usepackage{wrapfig,enumerate}
\usepackage{amsfonts}
\usepackage{amsmath}
\usepackage{slashed}
\usepackage{amssymb}
\usepackage{latexsym}
\usepackage{multirow}
\usepackage{hyperref}
\hypersetup{colorlinks=true,citecolor=red,linkcolor=magenta,urlcolor=blue}
\usepackage{relsize}
\usepackage{booktabs}
\usepackage{subfigure}
\usepackage{cleveref}
\usepackage{fancyhdr}
\usepackage{float}
\usepackage{graphicx}
\usepackage{microtype}
\usepackage{tabularx}
\usepackage{xcolor}
\usepackage{adjustbox,bm}
\usepackage[bottom]{footmisc}
\usepackage{orcidlink}
\usepackage{physics}
\newcolumntype{C}{>{$}c<{$}}
\newcolumntype{L}{>{$}l<{$}}
\newcolumntype{R}{>{$}r<{$}}

\title{$Z$ and Higgs Factory Implications of Two Higgs Doublets with First-Order Phase Transitions}
\preprint{KA-TP-17-2025, DESY-25-085, COMETA-2026-35}
\author[a,b]{Anisha\orcidlink{0000-0002-5294-3786},}
\author[c]{Francisco Arco\orcidlink{0000-0003-3651-1788},}  
\author[a]{Stefano Di Noi\orcidlink{0000-0002-1140-4073},}
\author[d,e]{Christoph Englert\orcidlink{0000-0003-2201-0667},} 
\author[a]{and Margarete M\"uhlleitner\orcidlink{0000-0003-3922-0281}}
\affiliation[a]{  Institute for Theoretical Physics, Karlsruhe Institute of Technology (KIT),
Wolfgang-Gaede Straße 1, D-76131, Karlsruhe, Germany
}
\affiliation[b]{Institute for Astroparticle Physics, Karlsruhe Institute of Technology,
Hermann-von-Helmholtz-Platz 1, D-76344, Eggenstein-Leopoldshafen, Germany
}
\affiliation[c]{Deutsches Elektronen-Synchrotron DESY,
Notkestr. 85, D-22607 Hamburg, Germany}
\affiliation[d]{School of Physics and Astronomy, University of Glasgow, 
University Avenue, Glasgow G12 8QQ, United Kingdom }
\affiliation[e]{Department of Physics and Astronomy, University of Manchester, Oxford Road, Manchester M13 9PL, United Kingdom}
\emailAdd{anisha@kit.edu}
\emailAdd{francisco.arco@desy.de}
\emailAdd{stefano.dinoi@kit.edu}
\emailAdd{christoph.englert@manchester.ac.uk}
\emailAdd{margarete.muehlleitner@kit.edu}

\abstract{We investigate the potential of future electron-positron colliders, such as FCC-ee and CEPC, to probe 2-Higgs-doublet models (2HDMs) that facilitate a strong first-order electroweak phase transition (SFOEWPT), a necessary condition for electroweak baryogenesis. Focusing on a 2HDM in the CP-conserving limit, we identify parameter regions consistent with an SFOEWPT and evaluate their compatibility with projected precision electroweak and Higgs measurements, as well as searches for exotic Higgs bosons. We show that radiative corrections to $e^+e^-\to hZ$ production introduce deviations in the cross section that are resolvable with the anticipated sub-percent precision at lepton colliders even when experimental outcomes of the LHC and $Z$ pole measurements are in agreement with the SM. This underscores the opportunities of a precision lepton collider to explore BSM quantum corrections to the Higgs sector more broadly.}
\begin{document}
%%%%%%%%%%%%%%%%%%%
\maketitle
\flushbottom
\allowdisplaybreaks
%%%%%%%%%%%%%%%%%%%
\section{Introduction}
\label{sec:intro}
%%%%%%%%%%%%%%%%%%%
The search for new physics beyond the Standard Model (SM) remains a high-value target of the present and future data-taking runs at the Large Hadron Collider (LHC), ramping up statistics towards its high-luminosity (HL) phase. Although no conclusive signs of new interactions have been observed so far, it is conceivable that new phenomena can manifest themselves in small departures of SM-expected coupling patterns correlated with new particle observations at large energies. Prime examples of such scenarios are provided by 2-Higgs-Doublet Models (2HDMs)~\cite{Lee:1973iz,Gunion:1989we,Branco:2011iw}. With experimental sensitivity increasing, particularly non-standard 2HDM scenarios~\cite{Anisha:2022hgv,Anisha:2023vvu} offer a dynamical explanation for the apparent absence of new physics signals at the LHC thus far. This perspective presents a theoretically compelling pathway to link established aspects of beyond-the-SM (BSM) physics with the experimental reach of the HL-LHC. Among these features is the requirement for a strong first-order electroweak phase transition (SFOEWPT) in the context of electroweak baryogenesis. While the SM in principle meets all three Sakharov conditions~\cite{Sakharov:1967dj} required for the dynamical generation of the observed baryon asymmetry in the universe, the 125~GeV Higgs boson implies a smooth cross-over~\cite{Kajantie:1996mn,Csikor:1998eu}, thus calling for a BSM Higgs sector extension.

The HL-LHC will provide tight constraints on the presence of new particles, should consistency with the SM prevail; the 2HDMs are no exception, and the experimental programme to achieve this is well underway. The phenomenological focus of new physics explorations will then shift towards the next generation of collider experiments, most likely precision $Z$ and Higgs boson factories such as FCC-ee~\cite{EuropeanStrategyGroup:2020pow} or CEPC~\cite{CEPCStudyGroup:2023quu}. With an abundant production of $Z$ bosons (the so-called GigaZ/TeraZ options) and $Z$-associated Higgs bosons, these environments enable the exploration of a relatively limited set of weak observables in comparison to hadron machines, however, with formidable precision. Such an experiment unfolds its true power when measurements are contrasted with theoretically motivated model correlations that shape the phenomenology of the TeV scale at the per cent level. In contrast, the plethora of ad-hoc interactions in effective field theory extensions such as the currently best-motivated SM Effective Field Theory (EFT) can imply blind directions~\cite{Chala:2025utt}, in particular when analysed in concert with radiative corrections~\cite{Asteriadis:2024xuk,Asteriadis:2024xts}. Whilst this does not equate to a breakdown of EFT methods, a na\"ive (often marginalised) treatment of independent interactions significantly waters down sensitivity prospects obtained by the relatively few available observables accessible at, e.g., an FCC-ee in comparison with hadron~machines.

In this work, we examine the potential of a precision $Z$ and $e^+ e^- \to hZ$ programme to inform a BSM-motivated parameter region of the 2HDM, namely the parameter region in which sufficient latent heat released during an SFOEWPT is available to meet the out-of-equilibrium requirement for electroweak baryogenesis as part of Sakharov's criteria~\cite{Sakharov:1967dj} (swiftly reviewed in Sec.~\ref{sec:elwba}). In this way, the 2HDM acts as a suitable scenario whose renormalisable correlations enable a highly quantitative analysis of the precision the FCC-ee will provide. To this end, we focus on the 2HDM of type I, and will firstly turn to the oblique corrections that drive its $Z$ pole programme in the absence of an LHC discovery in Sec.~\ref{sec:oblique}. There, we will also provide context with HL-LHC measurements of the 125 GeV Higgs particle. Further, in Sec.~\ref{sec:lhc}, we comment on the search for additional uncharged Higgs bosons at the HL-LHC that will inform the parameter space of a future $e^+e^-$ machine in parallel. In Sec.~\ref{sec:ee}, we discuss the radiative corrections to $e^+e^- \to hZ$ production in the SFOEWPT-preferred parameter region of the considered 2HDM model. We conclude in Sec.~\ref{sec:conc}.

%%%%%%%%%%%%%%%%%%%
\section{First-Order Electroweak Phase Transition in the 2HDM}
\label{sec:elwba}
%%%%%%%%%%%%%%%%%%%
Extended Higgs sectors as a promising avenue for electroweak baryogenesis have a long history. Whilst a range of precise understanding of bubble dynamics and their relation with efficient seeding of baryons remains an active area of research, in this work, we will impose a strong first-order phase transition via the criterion
\begin{equation}
\label{eq:criter}
    \xi_p\equiv\frac{v_p\!\left(T_p\right)}{T_p}>1\,.
\end{equation}
This relates the vacuum expectation value of the Higgs field at the percolation stage,  $v_p$, to the percolation temperature $T_p$. The parameter region characterised by such quantities can be considered sufficiently protected against baryon number washout at a stage in the thermal history of the universe when around a third of the comoving volume has been converted to the broken electroweak phase.\footnote{For an overview of related literature, cf.~e.g.~the recent review on the relation between particle physics, SFOEWPTs and gravitational waves \cite{Athron:2023xlk}.} We note that perturbative studies for scenarios with values of $\xi_p$ that are closer to the regime where the phase transition is not first order any more, can be misleading in studies of SFOEWPTs from extended scalar sectors, see, e.g.~\cite{Niemi:2024axp}. Our results should therefore be understood with these caveats in mind.

In this work, we consider the CP-conserving 2HDM~\cite{Lee:1973iz,Gunion:1989we,Branco:2011iw} as a theoretical framework to explore the existence of an SFOEWPT in BSM Higgs extended models. The 2HDM consists of the addition of a second $SU(2)_L$ complex Higgs doublet to the SM particle content. This model predicts the existence of five physical Higgs bosons: two neutral CP-even bosons $h$ and $H$ (with $m_h<m_H$), one neutral CP-odd boson $A$, and two charged bosons $H^\pm$. We define the mixing angle $\alpha$ as the angle that diagonalises the CP-even sector of the 2HDM, while the mixing angle $\beta$ diagonalises the CP-odd and charged sectors of the model. The modification factors of the couplings of the Higgs bosons to the gauge bosons with respect to the SM prediction, denoted by $\zeta_V^{h,H,A}$, are given by
\begin{equation}
    \zeta_V^h=\sin\!\left(\beta-\alpha\right)\,,\quad 
    \zeta_V^H=\cos\!\left(\beta-\alpha\right)\,,\quad 
    \zeta_V^A=0\,.
    \label{eq:zetabosons}
\end{equation}
We furthermore consider a discrete $\mathbb{Z}_2$ symmetry to avoid flavor-changing neutral currents~\cite{Glashow:1976nt,Paschos:1976ay}, such that $\phi_1\to\phi_1$ and $\phi_2\to-\phi_2$, where $\phi_{1,2}$ are the two Higgs doublets of the 2HDM. This symmetry is softly broken by the term $m_{12}^2\left(\phi_1^\dagger\phi_2 +\phi_2^\dagger\phi_1\right)$ in the Higgs potential. For convenience, we use the parameter ${\bar{m}^2}$, defined as 
\begin{equation}
\bar m^2 \equiv \frac{m_{12}^2}{\sin\beta\cos\beta}\,.
\end{equation}
In this work, we focus on the 2HDM type I, where the $\mathbb{Z}_2$ parities of the fermions are chosen so that they couple only to one of the Higgs doublets. This makes the Yukawa coupling modifiers universal among quarks and leptons. The coupling modifiers of the Yukawa interactions between fermions and Higgs bosons with respect to the SM are given by\footnote{The couplings to the pseudoscalar Higgs boson include an additional $i\gamma^5$.} \begin{equation}
\begin{gathered}
    \zeta_f^h = \sin\!\left(\beta-\alpha\right)+\cos\!\left(\beta-\alpha\right)\,\cot\beta \,, \\
    \zeta_f^H = \cos\!\left(\beta-\alpha\right)-\sin\!\left(\beta-\alpha\right)\,\cot\beta \,,\\
    \zeta_{u}^A=-\zeta_{d,l}^A= \cot\beta \,.
    \label{eq:zetafermions}
\end{gathered}
\end{equation}
In the following, we will use the following set of input parameters of the 2HDM, also known as the ``physical basis'',
\begin{equation}
    v\,, \quad
    m_h\,, \quad
    m_H\,, \quad
    m_A\,, \quad
    m_{H^\pm}\,, \quad
    \tan\beta\,, \quad
    \cos\!\left(\beta-\alpha\right)\,, \quad
    \bar m\,,
    \label{eq:input}
\end{equation}
where $v\simeq 246\ \mathrm{GeV}$ is the SM vacuum expectation value (vev) of the SM.

Currently, all signal strength measurements of the discovered Higgs boson are consistent with the SM prediction within the experimental and theoretical uncertainties~\cite{ATLAS:2022vkf,CMS:2022dwd}. Therefore, any viable BSM model with an extended Higgs sector must be able to accommodate an SM-like Higgs boson. Within the 2HDM, assuming that the SM-like Higgs boson is identified with the state $h$, there are two possibilities to achieve this~\cite{Gunion:2002zf,Carena:2013ooa,Bernon:2015qea}. The first one is by considering the limit $\cos\!\left(\beta-\alpha\right)\to0$ (or analogously $\beta-\alpha\to\pi/2$), since, in this limit, the tree-level couplings of $h$ to the SM particles yield the SM prediction, i.e.~$\zeta_V^h=\zeta_f^h=1$ (see Eqs.~\eqref{eq:zetabosons} and \eqref{eq:zetafermions}).\footnote{In this limit, the $h$ triple and quartic self-couplings also have the SM value at tree level.} This is often referred to as {\it alignment without decoupling}, or simply {\it alignment limit}, since the masses of the other BSM Higgs bosons can, in general, be light. The second option is to make the masses of the unobserved Higgs bosons very heavy, namely $m_H,\, m_A,\, m_{H^\pm}\gg v$, while keeping the quartic couplings in the potential at $\order{1}$. With these considerations, the alignment limit is obtained ($\cos\!\left(\beta-\alpha\right)\to0$) together with the condition that $\bar m\sim m_H,\, m_A,\, m_{H^\pm} \gg v$. This ensures that the scalar interactions between Higgs bosons do not increase with their masses, keeping their contributions to physical observables negligible. This is known as {\it alignment with decoupling}, or alternatively {\it decoupling limit}. While in the alignment limit, BSM bosons could, in principle, induce sizable radiative corrections (or, in other words, give rise to non-decoupling effects), in the decoupling limit, these corrections vanish, and the SM predictions are always recovered. This distinction between alignment and decoupling is crucial because the existence of an SFOEWPT in the 2HDM requires large radiative corrections in order to induce a potential barrier, and therefore it is incompatible with the decoupling limit (see, for instance, {Refs.~\cite{Dorsch:2014qja,Dorsch:2016nrg,Basler:2016obg,Dorsch:2017nza,Goncalves:2021egx,Goncalves:2022wbp,Biekotter:2023eil,Bittar:2025lcr} and references therein). 

To study the cosmological history of the 2HDM and determine whether an SFOEWPT took place, it is necessary to compute the effective potential at finite temperature given by
\begin{equation}
    V_\mathrm{eff}\!\left(T\right) = V_\mathrm{tree}\!\left(T=0\right) + V_\mathrm{CW}\!\left(T=0\right) + V_\mathrm{CT}\!\left(T=0\right)+V_\mathrm{T}\!\left(T\right) + V_\mathrm{daisy}\!\left(T\right)\,.
\end{equation}
We employ the public code \texttt{BSMPTv3}~\cite{Basler:2024aaf} to compute the above potential, which includes, at zero temperature, the tree-level potential $V_\mathrm{tree}$ of the 2HDM, the one-loop Coleman-Weinberg effective potential $V_\mathrm{CW}$~\cite{Coleman:1973jx}, and a finite counterterm potential $V_\mathrm{CT}$, such that an on-shell-like scheme is adopted to fix the mixing matrix elements in the Higgs mass matrices to their tree-level values at zero temperature \cite{Basler:2018cwe}. In consequence, the one-loop corrected input parameters in Eq.~\eqref{eq:input} are fixed to their tree-level values at $T=0$. In addition, the thermal corrections to the effective potential $V_\mathrm{T}$ are included in the high-temperature limit~\cite{Quiros:1994dr,Quiros:1999jp}. The temperature-corrected daisy resummation $V_\mathrm{daisy}$ is also included following the Arnold-Espinosa method~\cite{Arnold:1992rz}. For more details, we refer the reader to Refs.~\cite{Basler:2018cwe,Basler:2024aaf}.

To explore the possibility of an SFOEWPT within the 2HDM, we have performed a parameter scan of the model, such that the input parameters are varied randomly in the following intervals
\begin{equation}
\begin{gathered}
    m_H \in \left[150,\, 1500\right] \ \mathrm{GeV}\,, \quad 
    m_A,\ m_{H^\pm} \in \left[20,\, 1500\right] \ \mathrm{GeV}\,, \quad \\
    \tan\beta \in \left[0.5,\, 50\right]\,, \quad 
    \cos\!\left(\beta-\alpha\right) \in \left[-0.35,\, 0.35\right]\,, \quad 
    \bar m \in \left[0,\, 1500\right]\,,
\end{gathered}
\end{equation}
while $m_h=125.25$~GeV as it is identified with the SM-like Higgs boson. We confront the scan points with the most relevant theoretical and experimental constraints to date as implemented by the public code~{\tt{ScannerS}}~\cite{Coimbra:2013qq,Muhlleitner:2020wwk}. More concretely, on the theoretical side, we consider a point to be allowed if it satisfies the requirement of tree-level perturbative unitarity~\cite{Akeroyd:2000wc,Ginzburg:2005dt}, the potential is bounded from below (at $T=0$)~\cite{Deshpande:1977rw}, and if the EW minimum of the potential is the absolute minimum of the potential~\cite{Barroso:2013awa}. From the experimental side, all allowed points are required to lie in the $2\sigma$ uncertainty band of the {\it current} determination of the oblique parameters $S,\, T,\,U$~\cite{Peskin:1990zt,Grinstein:1991cd,Altarelli:1991fk,Peskin:1991sw,Burgess:1993vc} from the EW fit~\cite{Haller:2018nnx} at the one-loop level (using the formulae from~\cite{Grimus:2008nb}). Furthermore, we check that all points are allowed by the current bounds from LHC searches for BSM scalars and that the predicted SM-like Higgs boson signal strengths are within $2\sigma$ with respect to the SM prediction
(that is $\chi^2 - \chi^2_{\mathrm{SM}}<6.18$),%
\footnote{

This choice is widely used in the literature due to its convenience, since it does not require knowledge of the best fit in a generic model and it facilitates the comparison with other models.
This assumes that the SM is close to the best fit of the model (which is reasonable since we have not measured sizable deviations from the SM predictions) and that only the angles $\alpha$ and $\beta$ play an important role in the signal strength predictions.

}
where we apply these constraints with the help of the public code \texttt{HiggsTools}~\cite{Bahl:2022igd} (formerly known as \texttt{HiggsBounds}~\cite{Bechtle:2013wla,Bechtle:2011sb,Bechtle:2008jh} and \texttt{HiggsSignals}~\cite{Bechtle:2020uwn,Bechtle:2013xfa}). Finally, we check that the points are allowed at the 95\% CL level by present measurements of flavour-changing processes mediated by neutral currents~\cite{Haller:2018nnx}. In order to increase the amount of points allowed after an electroweak precision program at the $Z$ pole (as we will discuss in Sec.~\ref{sec:oblique}), we performed an additional dedicated scan with $m_A\sim m_{H^\pm}$ and/or $m_H\sim m_{H^\pm}$ to ensure that the contributions to the $T$ parameter (or alternatively the $\rho$ parameter) remain small~\cite{Pomarol:1993mu,Haber:2010bw}.

After applying all the above constraints, we compute the transition history of the allowed points in our scan with \texttt{BSMPT}. As discussed at the beginning of this section, we consider that a parameter point undergoes an SFOEWPT if it satisfies the condition in Eq.~\eqref{eq:criter}. For these points, we additionally demand that their potential is stable at next-to-leading order, and that the final phase corresponds to the broken electroweak phase with the SM vev at $T=0$.

%%%%%%%%%%%%%%%%%%%
\section{Phenomenology at the LHC and $e^+e^-$ Colliders}
%%%%%%%%%%%%%%%%%%%
The future collider roadmap is currently evolving in response to the Snowmass and European Strategy Update~\cite{Abidi:2025dfw}, yet a likely next step in terrestrial large-scale particle physics experiments is an electron-positron collider with staged precision $Z$ and Higgs programmes. Such environments will obviously draw from the insights of the HL-LHC findings that could well pinpoint new physics through small yet relevant coupling modifications.

However, to highlight the true sensitivity uplift that an integrated $e^+e^-$ programme can achieve, specifically in the context of concrete UV-motivated scenarios, we will adopt a somewhat pessimistic outlook in this section, namely that the LHC at the end of its HL phase will not provide conclusive evidence for new physics. On the one hand, for the example of our 2HDM type I scenario, we will gather evidence that this could be possible even if the new scale of physics is relatively low and well within the kinematic reach of the LHC. On the other hand, we will show that within the assumptions of the model that we focus on in this work, a precision $hZ$ programme will typically lead to discovery.

%%%%%%%%%%%%%%%%%%%
\subsection{SFOEWPT versus a Precision $Z$ Pole Programme }
\label{sec:oblique}
%%%%%%%%%%%%%%%%%%%
As the first step in our study, we tension the parameter region identified in Sec.~\ref{sec:elwba} against an electroweak precision $Z$ pole programme, i.e.\ the first stage of a lepton collider such as FCC-ee or CEPC (see also~\cite{Aiko:2020ksl}). In particular, oblique corrections~\cite{Peskin:1990zt,Grinstein:1991cd,Altarelli:1991fk,Peskin:1991sw,Burgess:1993vc} can impose significant constraints on the 2HDM parameter space due to non-SM gauge dynamics. The $1\sigma$ projected sensitivity for the $S$ and $T$ oblique parameters for the FCC-ee is~\cite{deBlas:2019rxi,EuropeanStrategyforParticlePhysicsPreparatoryGroup:2019qin}
\begin{equation}
    \sigma_S=0.0038\,, \quad
    \sigma_T=0.0022\,, \quad
    \rho_{ST}=0.724\,,
\end{equation}
where $\rho_{ST}$ is the correlation coefficient between $S$ and $T$.

%%%%%%%%%%%%%%%%%%%
\begin{figure}[!t]
    \subfigure[\label{fig:zst}]{\includegraphics[width=0.49\textwidth]{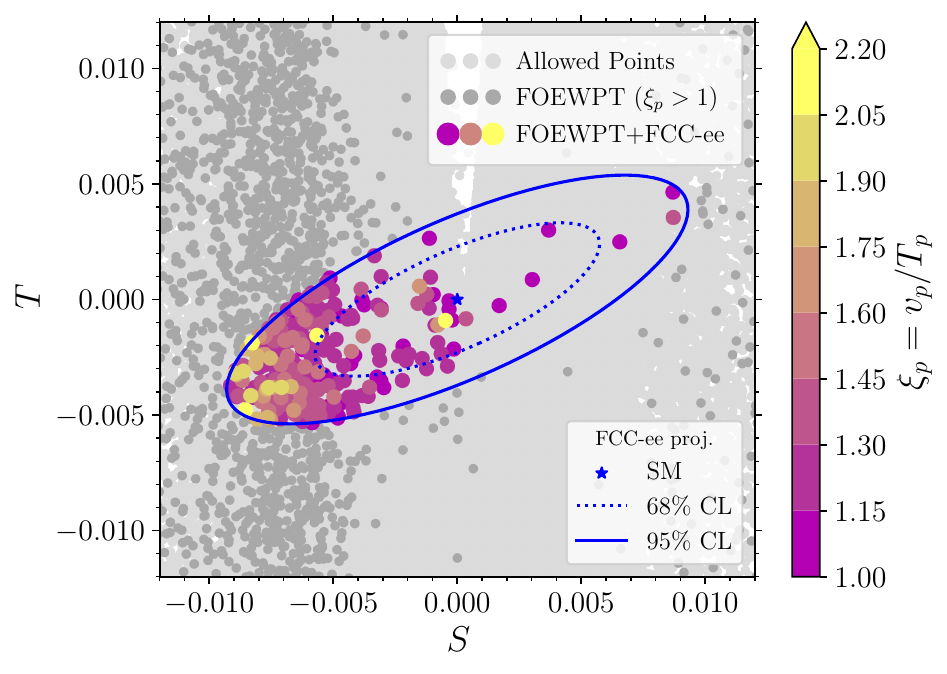}}\hfill
    \subfigure[\label{fig:zmass}]{\includegraphics[width=0.49\textwidth]{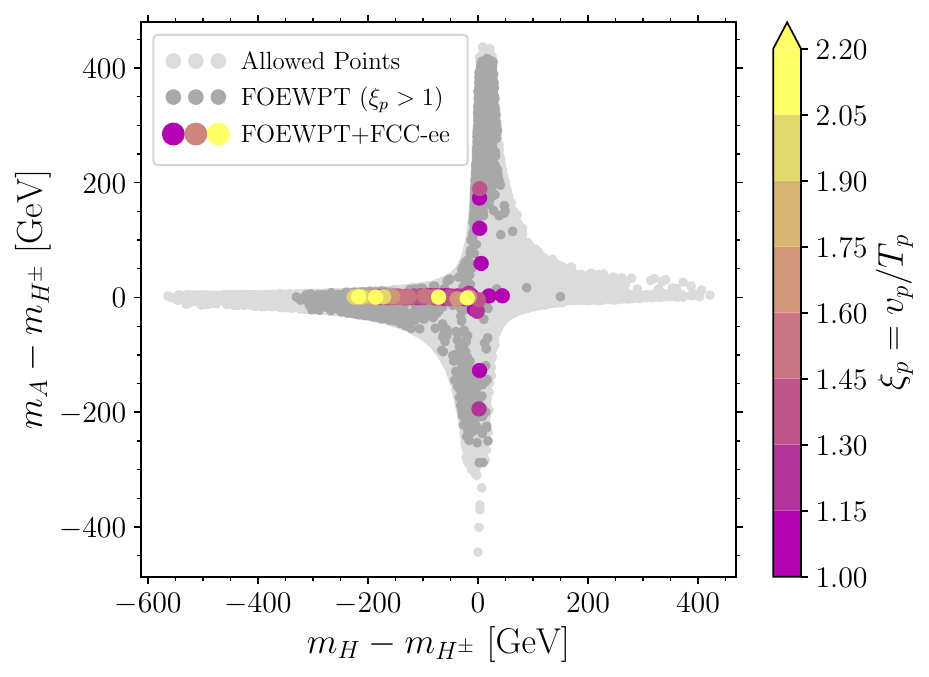}}
    \subfigure[\label{fig:zm}]{\includegraphics[width=0.49\textwidth]{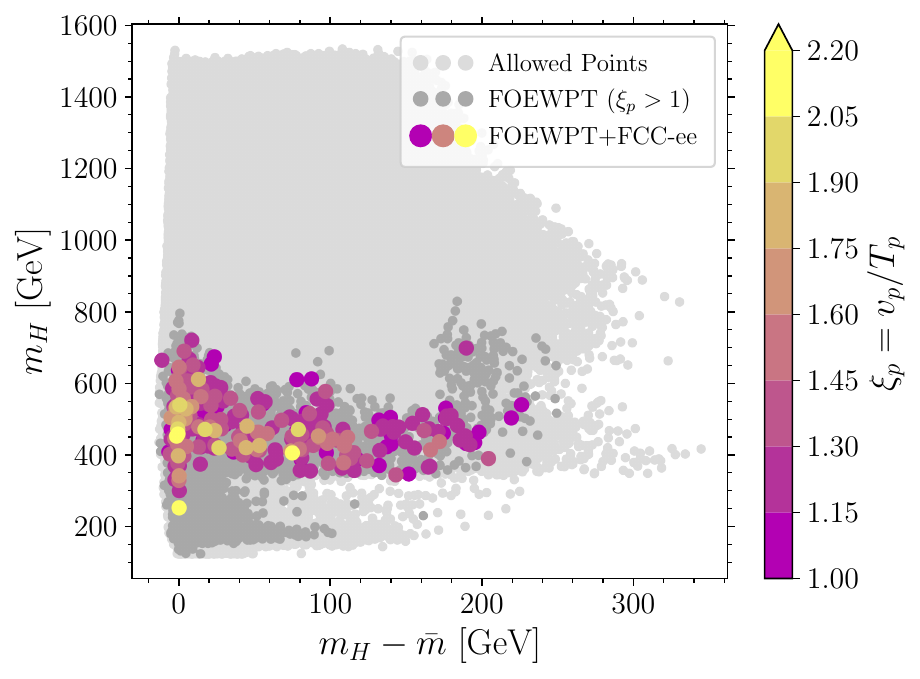}}\hfill
    \subfigure[\label{fig:zlhc}]{\includegraphics[width=0.49\textwidth]{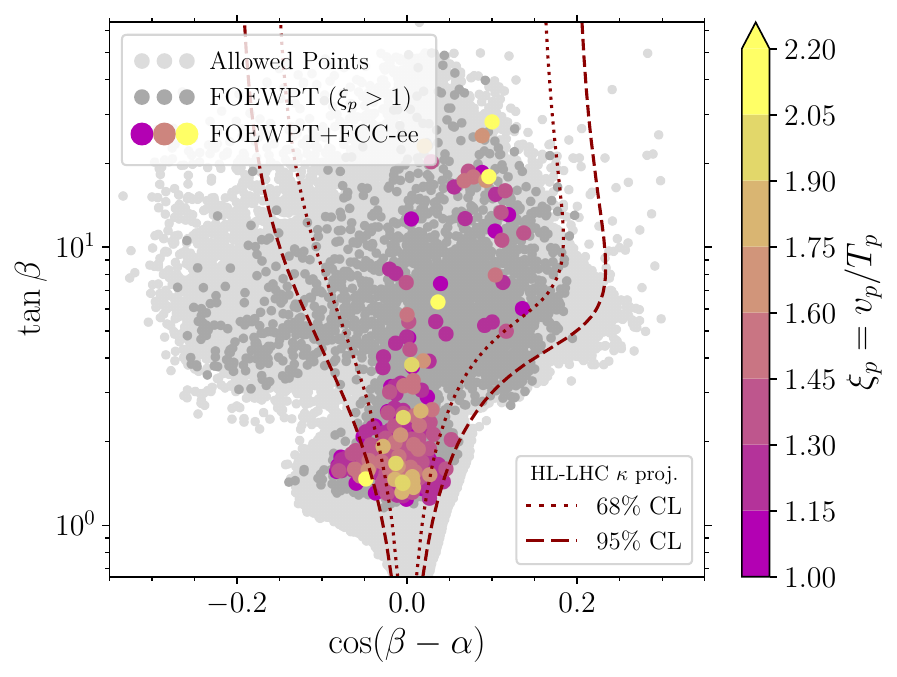}}
    \caption{\label{fig:plotZ} Projections of the SFOEWPT scan as detailed in the main body of the text. Panel~\ref{fig:zst} shows the compatibility with the $S$ and $T$ parameters at 68\% CL (dotted blue line) and 95\% CL (full blue line). Here, light grey points are parameter points compatible with the constraints specified in Sec.~\ref{sec:elwba}, dark grey points additionally fulfil the SFOEWPT constraint $\xi_p >1$, and colored points fulfil the projected constraints on the EWPOs representatively obtained at a future FCC-ee for its $\sqrt{s}=m_Z$ phase. The $Z$-pole programme will efficiently single out a relatively small parameter space of the SFOEWPT-preferred parameter region. Panels~\ref{fig:zmass} and \ref{fig:zm} show the contours of the allowed mass spectra after successively applying the constraints. Panel~\ref{fig:zlhc} highlights the relation of this region in the context of the expected 125 GeV signal strength constraints obtainable at the HL-LHC at 68\% CL (dotted red line) and 95\% CL (dashed red line), displayed in the $\tan\beta$ versus $\cos(\beta-\alpha)$ plane. }
\end{figure}
%%%%%%%%%%%%%%%%%%%

The results for our parameter scan are presented in Fig.~\ref{fig:plotZ}. As expected, parameter choices that are relatively far from the alignment limit can be efficiently constrained with a precision $Z$ pole programme at an $e^+e^-$ machine. Sensitivity here extends well beyond the sensitivity of a coupling analysis for the 125 GeV Higgs boson at the HL-LHC, which is highlighted in Fig.~\ref{fig:zlhc}.\footnote{To obtain the confidence intervals of the HL-LHC we performed a $\chi^2$ fit to test the projected values of the coupling modifiers to $b$- and $t$-quarks, $W$ and $Z$ bosons, gluons, and $\tau$ and $\mu$ leptons from Ref.~\cite{CMS:2025hfp} againts the 2HDM predictions from Eqs.~\eqref{eq:zetabosons} and~\eqref{eq:zetafermions}.} The requirement of a significant mass splitting between the heavy neutral scalar and pseudoscalar mass, $m_H$ and $m_A$, respectively, to achieve an SFOEWPT in the 2HDM (see also~\cite{Dorsch:2014qja,Dorsch:2016nrg,Basler:2016obg,Dorsch:2017nza,Biekotter:2023eil}) stands in direct tension with an SM-like outcome of the oblique electroweak precision observables (EWPO) analysis for the expected FCC-ee precision, cf.~Fig.~\ref{fig:zmass}. Points that still admit an SFOEWPT according to Eq.~\eqref{eq:criter} are forced towards the alignment limit $\cos(\beta-\alpha)\simeq 0$, yet at relatively small mass scales, to modify the electroweak cross-over to an SFOEWPT, cf.~Fig.~\ref{fig:zm}. As already mentioned, signal strength analyses at the HL-LHC do not provide significant constraints for this parameter range. However, the mass scale of the exotic Higgs bosons falls into a range that is well accessible by ATLAS and CMS. We will turn to this in the next section.

%%%%%%%%%%%%%%%%%%%
\subsection{Context with LHC Higgs Partner Searches}
\label{sec:lhc}
%%%%%%%%%%%%%%%%%%%
If the exotic states have masses above the $t\bar{t}$ threshold, the heavy scalars $H$ and $A$ predominantly decay into these final states. This holds for the largest part of the successful parameter points of our scan, detailed in Sec.~\ref{sec:elwba}. Both ATLAS and CMS are actively searching for heavy scalars in the $t\bar{t}$ channels~\cite{CMS:2019pzc,ATLAS:2024vxm,CMS:2025dzq}. It is well-established that interference effects, both between different Higgs resonances (signal–signal interference)~\cite{Basler:2019nas,Bahl:2025you} and between the Higgs signal and the QCD background, can significantly distort the expected Breit–Wigner resonance shapes~\cite{Gaemers:1984sj}. In some cases, these effects can even result in a complete cancellation of the signal~\cite{Jung:2015gta}, or severely reduce experimental sensitivity, particularly when the net effect is a depletion rather than an excess of events.

Of course, other search strategies for BSM states exist.
Examples are, e.g., $A\to ZH$ or $H\to ZA$ searches, which are motivated by the large mass splitting typically observed between $A$ and $H$ in the SFOEWPT favoured region~\cite{Biekotter:2023eil,Dorsch:2014qja,Basler:2016obg}. Other BSM searches include charged Higgs production channels (for a recent sensitivity extrapolation to the HL-LHC phase see~\cite{Atkinson:2022pcn}), and four-top quark production, which can mitigate the interference distortion despite the small production rate~\cite{Anisha:2023xmh}. However, these searches generally yield a smaller production cross section compared to gluon fusion production of neutral scalars and their subsequent decay into top-quark pairs. Therefore, we will focus on the latter here (we will come back to the LHC discovery potential for the 2HDM more broadly further below).

%%%%%%%%%%%%%%%%%%%
\begin{figure}[!t]
\centering
\includegraphics[width=0.52\textwidth]{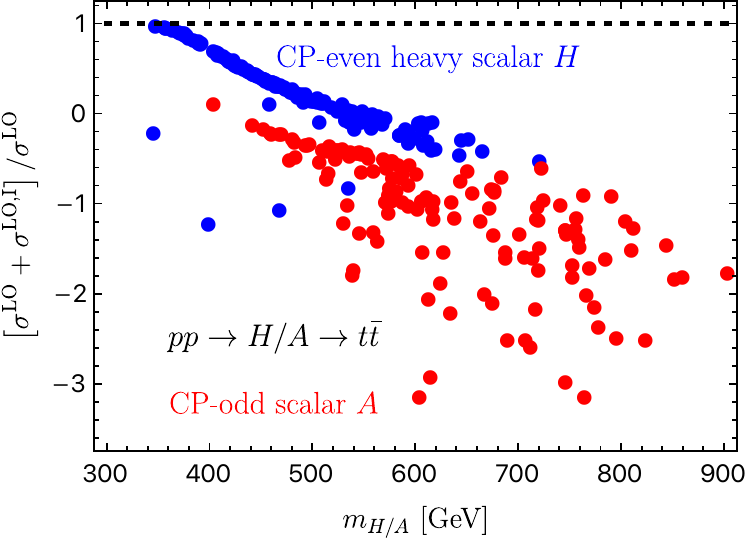}
\caption{\label{fig:pptt} Interference-corrected (signal-signal as well as signal-background) cross sections compared to the pure signal process for LHC production of the uncharged heavy scalars, $H$ (blue points) and $A$ (red points), respectively, as a function of their respective mass, for the parameter points of the scan detailed in Sec.~\ref{sec:elwba}.}
\end{figure} 
%%%%%%%%%%%%%%%%%%%

In Fig.~\ref{fig:pptt}, we illustrate the combined effect of signal-signal and signal-background interference for our parameter scan by comparing the interference-corrected cross section with the leading-order approximation. To this end we define the signal and the background amplitudes ${\cal{M}}_S, \, {\cal{M}}_\text{rest}$, where `rest' includes QCD ($gg\to t\bar t$ at ${\cal{O}}(\alpha_s)$) and Higgs contributions of the remaining Higgs bosons (e.g.~$h,A$ in case of $H$ production). The signal cross section is readily obtained as 
\begin{equation}
\hbox{d} \sigma^{\text{\text{LO}}} \sim |{\cal{M}}_S|^2
\end{equation}
and displays the usual Breit-Wigner behaviour in the $m_{t\bar t}$ invariant mass. For definiteness, we sample cross sections around the exotic scalar $S$ with $m_{t\bar t} \in [m_S-5\,\Gamma_S,m_S+5\,\Gamma_S]$. The interference contribution is given by 
\begin{equation}
\hbox{d} \sigma^{\text{\text{LO,I}}} \sim 2 \text{Re} \left( {\cal{M}}_S^\ast {\cal{M}}_\text{rest}\right).
\end{equation}

For the parameter points sampled in our scan of Sec.~\ref{sec:elwba}, we observe destructive interferences, which can erase the net cross section when integrated across the scalar threshold or lead to an underproduction due to interference with the QCD background. Whilst the experiments are taking into account these effects, they are intrinsically model-dependent, and given the results of Fig.~\ref{fig:pptt}, it is conceivable that the LHC might not be able to fully exclude the parameter range that is highlighted by our scan region, although it is kinematically accessible.

We stress that the results presented here should not be understood as a `no-go theorem' for the discovery of such states at the LHC. In fact, both ATLAS and CMS~\cite{CMS:2025dzq,ATLAS:2024vxm,CMS:2019pzc} incorporate interference effects in their likelihood analyses, yet with simplifying assumptions that make a direct comparison with our 2HDM scenario opaque. For instance, the recent ATLAS analysis~\cite{ATLAS:2024vxm} excludes parameter regions in the 2HDM type II with $m_H=m_A>300~\text{GeV}$ with $\tan\beta\lesssim 1.5-3$. It is worth highlighting that these exclusions are severely impacted by systematic uncertainties, as demonstrated by ATLAS. 

The results for a 2HDM type I (as considered in this paper) would arguably be similar to those reported by ATLAS, because the phenomenology of all 2HDM types is very similar for $\tan\beta\sim1$, and, moreover, the Higgs boson coupling to the top quark is universal for all types.
Therefore, the ATLAS findings serve as a strong indicator that the LHC {\it{is}} gaining sensitivity in the mass region we consider.\footnote{The observation of an excess near the $t\bar t$ threshold consistent with a ``toponium'' bound state by both ATLAS~\cite{ATLAS:2025kvb} and CMS~\cite{CMS:2025kzt} is evidence of this.
However, precise non-relativistic QCD predictions of such a bound state will be necessary in the future for BSM searches in the $t\bar t$ channel in this mass range.
}
Especially towards the HL-LHC phase, we can therefore expect these analyses to become increasingly sensitive, also to dip structures in the $m_{t\bar t}$ spectrum, which could then facilitate a discovery when the interference becomes large. To our knowledge, no HL-LHC extrapolation is currently available, but such analyses remain high-priority lines of BSM investigations with the potential to reveal new physics. In light of this, we will assume an overly pessimistic view that the LHC will not fully explore this region to show that an $e^+e^-$ machine can close this gap.
Furthermore, other top-phillic search channels for neutral Higgs partners, such as top quark pair-associated and four top quark production, do not suffer from such sensitivity-altering effects. These, therefore, provide additional sensitivity; however, at a significantly reduced cross section and consequently limited sensitivity reach.

%%%%%%%%%%%%%%%%%%%
\subsection{SFOEWPT versus Precision Higgs-Associated Production}
\label{sec:ee}
%%%%%%%%%%%%%%%%%%%
\subsubsection{Elements of the Calculation}
\label{sec:eecalc}
%%%%%%%%%%%%%%%%%%%
Radiative corrections in the 2HDM at electron-positron machines have been considered in, e.g., Refs.~\cite{Lopez-Val:2010asi,Xie:2018yiv,Abouabid:2020eik,Aiko:2021nkb}. We organise our computation of the next-to-leading order (NLO) corrections to the Higgs-strahlung process $e^+ e^- \to hZ$ similar to~\cite{Lopez-Val:2010asi}, see Fig.~\ref{fig:feyntop} for its representation in terms of Feynman diagrams. The various one-loop vertex functions depicted in grey in Fig.~\ref{fig:feyntop} are computed using {\tt{FeynArts}}\footnote{We used the {\tt{FeynArts}} built-in model file {\tt{THDM.mod}} which is based on the 2HDM potential as parameterised in the {\it Higgs Hunter's Guide}~\cite{Gunion:1989we}.}/{\tt{FormCalc}}/{\tt{FeynCalc}}/{\tt{LoopTools}}~\cite{vanOldenborgh:1989wn,Mertig:1990an,Hahn:2000kx,Hahn:1998yk,Hahn:2000jm,Shtabovenko:2016sxi,Shtabovenko:2020gxv}. They are then contracted with the leading order topologies to obtain the expression ready for numerical~integration. 

%%%%%%%%%%%%%%%%%%%
\begin{figure}[!b]
\centering
\includegraphics[width=0.99\textwidth]{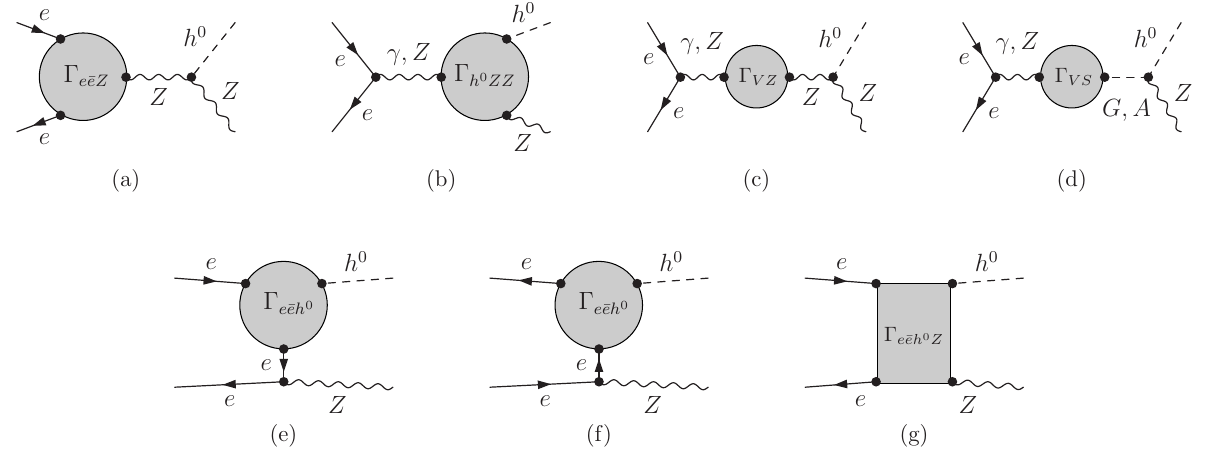}
\caption{\label{fig:feyntop} Feynman diagram toplogies contributing at NLO to $e^+ e^- \to hZ$ in the 2HDM. In particular, (b) and (d) are sensitive to Higgs-mixing effects. We omit $m_e\neq 0$ effects due to vanishingly small electron-Higgs couplings.}
\end{figure} 
%%%%%%%%%%%%%%%%%%%

A non-vanishing electron mass is employed for the gauge-invariant QED part to regularise the soft photon emission below a threshold $\Delta E$. This part can be cancelled analytically~\cite{tHooft:1978jhc,Denner:1991kt} against the corresponding real emission contribution according to the Bloch-Nordsiek~\cite{Bloch:1937pw} or Kinoshita-Lee-Nauenberg~\cite{Kinoshita:1962ur,Lee:1964is} theorems. As the real-emission contribution at the considered order in perturbation theory is insensitive to the virtual presence of the BSM mass spectrum, we do not include the contribution above $\Delta E$. Experimentally, this also corresponds to a different, resolvable final state, which should be investigated separately. In the following, we will conservatively choose $\Delta E=30~\text{GeV}$. At intermediate points of the calculation, IR singularities arising from QED radiative corrections~Fig.~\ref{fig:feyntop} (a) (soft photon singularities appear as part of the QED vertex correction) are regularised with a photon mass.\footnote{This operation is automatically handled in the scalar function implementation of \tt{LoopTools}.} The analytical form of the soft counterterms cancels this cut-off and replaces it with a real-emission related finite term as a function of $\Delta E$, proportional to the Born-level amplitude. We have checked this cancellation numerically. 

The remaining weak parts of the amplitude can be computed in the $m_e=0$ approximation, exploiting the numerical irrelevance of the electron mass. This allows us to neglect diagrams involving the Higgs Yukawa coupling to an electron-positron pair, simplifying the computation of the matrix element considerably. Cross-checks are performed with a second independent computation. Throughout, we use the Fleischer-Jegerlehner tadpole scheme~\cite{Fleischer:1980ub}, in which the bare tadpole is set to zero. This scheme has been implemented for the 2HDM and referred to as ``alternative tadpole scheme'' in Refs.~\cite{Krause:2016oke,Krause:2016xku,Krause:2017mal,Krause:2018wmo},\footnote{For further literature on the renormalisation of the 2HDM, cf.~\cite{Kanemura:2004mg,Denner:2016etu,Altenkamp:2017ldc,Altenkamp:2017kxk,Fox:2017hbw,Grimus:2018rte,Denner:2018opp,Dittmaier:2022maf,Dittmaier:2022ivi,Kanemura:2024ium,Guerandel:2025kjq}.} where in the two-point and three-point functions, the relevant tadpole contributions are included alongside the usual one-particle irreducible (1PI) diagrams. For the renormalisation of the 2HDM mixing angles, we follow the prescription of~\cite{Krause:2016oke,Krause:2016xku,Krause:2018wmo} and use the on-shell tadpole-pinched scheme, introduced there, based on evaluating the scalar two-point functions using the pinch technique~\cite{Binosi:2009qm}.\footnote{The pinch technique was adopted to extract the gauge-independent contributions to the mixing angle counterterms.} The angle counterterms are defined through the scalar off-diagonal wavefunction renormalisation constants in the alternative tadpole scheme with additional UV-finite parts evaluated at on-shell scalar boson masses. For the renormalisation of the fields and masses, using the on-shell renormalisation conditions, the field and mass renormalisation constants are obtained as in Ref.~\cite{Denner:1991kt}. The UV-finiteness after the implementation of the renormalisation programme has been checked analytically as well as numerically. The numerical integration has been compared against {\tt{MadGraph5\_aMC@NLO}}~\cite{Alwall:2014hca} for the SM. We have further checked that our 2HDM calculation approaches, at leading and at next-to-leading order, the SM prediction in the decoupling limit\footnote{We obtain the cross section in the decoupling limit by setting $m_H=m_A=m_{H^\pm}=\bar m=1.8\, \mathrm{TeV}$ and $\cos\!\left(\beta-\alpha\right)=0$, see discussion in Sec.~\ref{sec:elwba}. This result is robust against changes of the mass scale.} (we obtain the SM result with a second independent implementation).

%%%%%%%%%%%%%%%%%
\begin{figure}[!t]
\centering
\includegraphics[width=0.6\textwidth]{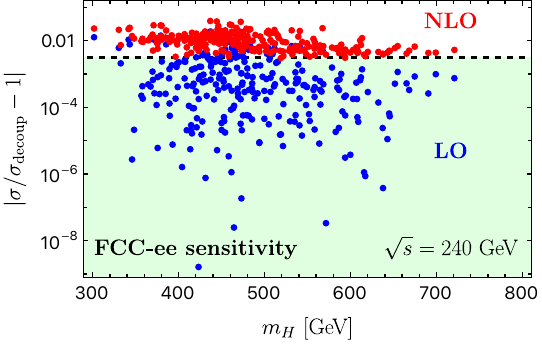}
\caption{\label{fig:eezh} Production of the SM-like Higgs boson in association with a $Z$ boson at the FCC-ee, in relation to the decoupling limit at $\sqrt{s}=240~\text{GeV}$ for the scan detailed in Sec.~\ref{sec:elwba}. Blue points denote leading-order predictions, which are all compatible with the expected FCC-ee sensitivity of $hZ$ production of $\Delta \sigma/\sigma =0.31\%$~\cite{Selvaggi:2025kmd}; all points are compatible with an SM outcome of the FCC-ee $Z$-pole programme. Including radiative corrections (red points) modifies these predictions, creating the discovery potential at a lepton collider even when $Z$ pole observables are insensitive. The green region is the experimental sensitivity $\Delta \sigma/\sigma =0.31\%$. The decoupling cross section $\sigma_{\text{decoup}}$ is evaluated consistently at LO (blue points) and NLO (red points).}
\end{figure} 
%%%%%%%%%%%%%%%%%%%

%%%%%%%%%%%%%%%%%%%
\subsubsection{Higgs Production in $e^+e^-$ Collisions - Discrimination Beyond EWPOs}
%%%%%%%%%%%%%%%%%%%
We are now ready to turn to the implications of our previous results for a precision investigation of associated Higgs production at an $e^+e^-$ collider. A Higgs machine will typically run at a centre-of-mass energy of $\sqrt{s}=240~\text{GeV}$ where the $e^+e^- \to hZ$ cross section has a maximum. The number of observables that are accessible in this instance is limited compared to hadron machines that probe a plethora of processes across a large range of energy scales. However, when viewed in the context of concrete model predictions, the overall sensitivity that can be expected at an FCC-ee of~\cite{Selvaggi:2025kmd}
\begin{equation}
\label{eq:eezh}
\frac{\Delta \sigma(e^+ e^- \to hZ)}{\sigma(e^+ e^- \to hZ) } = 0.31\%\,,
\end{equation}
predominantly achieved from an analysis of the $Z$ boson recoil spectrum~\cite{FCC:2018evy,Azzurri:2021nmy, TLEPDesignStudyWorkingGroup:2013myl}, carries a very large potential for new physics discovery.

The parameter region of our 2HDM scan that is consistent with the $Z$ pole programme is forced into the 2HDM alignment limit. Hence, it is not surprising that most of the parameter points are consistent with the uncertainty of Eq.~\eqref{eq:eezh} at the leading order. Consistency with $S,\,T$ forces the interactions of the SM-like Higgs boson into a region where the leading-order cross section is largely consistent with the SM (the 2HDM cross section in the {\it{decoupling limit}} is around $\sigma_{\mathrm{decoup}}=0.222$~pb at LO and $0.211$~pb at NLO), cf.~Fig.~\ref{fig:eezh}. 

The requirement of an SFOEWPT, however, is incompatible with this decoupling limit (the latter approaches the SM for which the phase transition is a cross-over). Hence, we can expect quantum corrections to be relevant for the points that single out an SFOEWPT, also when they are aligned with the SM expectation of $Z$ pole measurements. Indeed, as shown in Fig.~\ref{fig:eezh}, the weak radiative corrections as detailed in Sec.~\ref{sec:eecalc}, modify the 2HDM cross sections away from the decoupling limit (and hence, away from the SM). More importantly, the modifications are typically large enough for the $e^+e^-\to hZ$ programme to detect new physics based on the expectation of Eq.~\eqref{eq:eezh}.

It is worthwhile mentioning that the discrimination on the basis of the $hZ$ cross section is not in a one-to-one correspondence with the requirement of an SFOEWPT. It is the predictive power of renormalisability that drives experimentally detectable model correlations beyond leading order in a theoretically controlled way. An SFOEWPT is driven by the requirement of a relatively light exotic particle spectrum and a departure from SM-expected Higgs interactions. In any UV-complete scenario (and in particular for the 2HDM considered here), these requirements select a region of the parameter space that is, to a large extent, experimentally distinguishable from the SM in $hZ$ production at a 240 GeV lepton collider. The radiative corrections are not dominantly sensitive to the mass splittings themselves, but to their correlated effects in terms of inter-Higgs couplings and radiative corrections to mixing angles. These effects are two-loop suppressed for the $Z$ pole programme; $hZ$ production accessing these correlation modifications at one-loop level makes them large enough for detection within the expected experimental sensitivity. It is known that modifications of the Higgs self-coupling can be parametrically large~\cite{Kanemura:2004mg,Braathen:2019pxr,Arco:2022lai,Arco:2025pgx}, including regions with SFOEWPTs~\cite{Goncalves:2021egx,Dorsch:2017nza,Biekotter:2023eil,Basler:2017uxn}. These are formally two-loop suppressed effects, which could, however, be numerically relevant for the given sensitivity that can be achieved at a machine like FCC-ee. An analysis of such effects is only meaningful as part of a comprehensive two-loop investigation, which is beyond the scope of our present work.

%%%%%%%%%%%%%%%%%%%
\section{Summary and Conclusions}
\label{sec:conc}
%%%%%%%%%%%%%%%%%%%
In this work, we have considered an electroweak precision $Z$ and Higgs analysis programme in light of expected LHC sensitivities for Higgs measurements and searches for exotics. Based on the sole assumption of a strong first-order electroweak phase transition, we have clarified the discovery potential for new physics at each experimental stage when moving from the HL-LHC to $e^+e^-$ collisions at different energies.

The generic structure of our discussion has been deliberately (and, perhaps, see below, unnecessarily) pessimistic. We have mainly focused on parameter regions of SFOEWPTs that are compatible with an SM-consistent outcome of precision $Z$ pole measurements. In parallel, we argued that the most dominant exotic Higgs discovery modes potentially miss the mass scales imposed by the SFOEWPT. Nonetheless, the outcome of our analysis is extraordinarily encouraging. Most of the parameter points that fulfil these pessimistic criteria for the 2HDM type I considered in this work show a large enough cross section deviation in $hZ$ production at NLO for new physics to be discovered. The reason behind this is the congeniality of two observations: (i) the relevance of radiative corrections when BSM is aligned with but not decoupled from the SM, and (ii) the expected experimental sensitivity to $hZ$ production achievable at lepton colliders.

Of course, our observations are specific to the 2HDM that we have considered here. Nonetheless, we can expect our findings to generalise to other scenarios as well, qualitatively. Moving away from the SM to address (parts of) its shortcomings in the light of the current consistency of LHC results with SM expectations will leave sizeable quantum footprints that can be targeted at experimental precision environments. In this sense, our results are representative and have provided convincing evidence that such parameter regions can be efficiently probed by $e^+e^-\to hZ$ measurements, which are driven by relatively model-independent $Z$-recoil studies. Therefore, if the more profound theory that underpins the SM follows the behaviour of established quantum field theory, there is no a priori reason beyond tuning for it to be missed at the next generation of lepton colliders {\emph{at the latest}}.

Along this line, it is worthwhile pointing out that other $e^+e^-$ machines have been proposed, with or without a $Z$ pole programme, e.g.~ILC~\cite{Bambade:2019fyw}, CLIC~\cite{CLICdp:2018cto} or LCF~\cite{LinearColliderVision:2025hlt,LinearCollider:2025lya}. The precision of an $hZ$ measurement will be quantitatively similar. A less precise measurement of the electroweak input parameters that is provided by a dedicated $Z$ run can result in somewhat larger theoretical interpretation uncertainty. We have not considered this in detail, and within the approximations of our work, we can expect such a machine to have a similar sensitivity reach as the FCC-ee concept. Relaxing the $S,T,U$ constraints, which serve the purpose of aligning the Higgs couplings alongside the exotics' mass splitting, a more general phenomenological NLO pattern becomes possible. The additional parametric freedom can then also lead to NLO cross sections consistent with the SM uncertainty. Such a limitation can potentially be overcome by studying exclusive decay modes in $ Zh $ production, which provide a largely independent measurement of couplings and mixing angles as the Higgs boson is narrow.

Finally, it is worthwhile mentioning that the new physics as considered here can be discovered at {\emph{any}} stage in the present and future collider programme, e.g. through small coupling deviations of the SM-like Higgs boson at the HL-LHC in the near future (e.g. through improving beyond the single-Higgs sensitivity expectations with growing data sets~\cite{Belvedere:2024wzg} or through discovery in rare channels). We highlight again that the experiments are actively including interference effects in their exotics' searches for $gg\to H/A$, and we can expect analyses to fully reflect model correlations beyond benchmarks, especially when we turn to the HL-LHC phase, when backgrounds will become increasingly under statistical and systematic control. This also includes search channels that we have explicitly not considered in this work. In the lucky instance of new physics discovery, the combination of a $Z$ pole programme (for a more comprehensive analysis see Refs.~\cite{Chen:2018shg,Chen:2019pkq}) in tandem with an experimentally robust precision investigation of Higgs production at an electron-positron collider will lead to a detailed understanding of the nature of the electroweak scale. Whether or not an $e^+e^-$ collider remains a motivated concept will, of course, depend on the nature of the discovery, which could instead justify a direct consideration of FCC-hh.

%%%%%%%%%%%%%%%%%%%
\subsection*{Acknowledgements}
%%%%%%%%%%%%%%%%%%%
We thank J.~de Blas for helpful conversations regarding the expected precision of the $S,\ T$ fit at future lepton colliders. We also thank L.~Biermann and C.~Borschensky 
for helpful discussions.
A., S.D.N., M.M. acknowledge support by the Deutsche Forschungsgemeinschaft (DFG, German Research Foundation) under grant 396021762 - TRR~257.
F.A.~acknowledges support by the Deutsche Forschungsgemeinschaft (DFG, German Research Foundation) under Germany's Excellence Strategy -- EXC 2121 ``Quantum Universe'' -- 390833306. The work of F.A.~has also been partially funded by the Deutsche Forschungsgemeinschaft (DFG, German Research Foundation) -- 491245950.
C.E. is supported by the STFC under grant ST/X000605/1, and by the Leverhulme Trust under Research Fellowship RF-2024-300$\backslash$9. C.E. is further supported by the Institute for Particle Physics Phenomenology Associateship Scheme.
%
%%%%%%%%%%%%%%%%%%%
\bibliographystyle{JHEP}
\bibliography{references}
%%%%%%%%%%%%%%%%%%%
\end{document}